\documentclass[prl,twocolumn]{revtex4}
\usepackage{amsfonts}
\usepackage{amssymb}
\usepackage{newlfont}
\usepackage{stmaryrd}
\usepackage{mathrsfs}
\usepackage{euscript}
\usepackage{color}
\usepackage{graphicx}

\begin{document}

\newcommand{\pr}{\partial}
\newcommand{\rta}{\rightarrow}
\newcommand{\lta}{\leftarrow}
\newcommand{\ep}{\epsilon}
\newcommand{\ve}{\varepsilon}
\newcommand{\p}{\prime}
\newcommand{\om}{\omega}
\newcommand{\ra}{\rangle}
\newcommand{\la}{\langle}
\newcommand{\td}{\tilde}
\newcommand{\dg}{\dagger}
\newcommand{\mo}{\mathcal{O}}
\newcommand{\ml}{\mathcal{L}}
\newcommand{\mathp}{\mathcal{P}}
\newcommand{\mq}{\mathcal{Q}}
\newcommand{\ms}{\mathcal{S}}
\newcommand{\llra}{\longleftrightarrow}
\newcommand{\nl}{$\newline$}
\newcommand{\nll}{$\newline\newline$}

\title{Phase controllable dynamical localization: a generalization of the Dunlap-Kenkre result}

\author{Navinder Singh}

\email{navinder.phy@gmail.com}

\affiliation{Physical Research Laboratory, Navrangpura, Ahmedabad-380009 India.}

\begin{abstract}
Dunlap-Kenkre result states that Dynamical Localization (DL) of a field driven quantum particle in a discrete
periodic lattice happens when the ratio of the field magnitude to the field frequency (say, $\eta$) of the
diagonal sinusoidal drive is a root of the ordinary Bessel function of order 0. This has been
experimentally verified. A generalization of the Dunlap-Kenkre result is presented here. We analytically show
that if we have an off-diagonal driving field (with modulation $\delta$) and diagonal driving field with
different frequencies (say $\omega_1$ and $\omega_2$ respectively) and a definite phase relationship $\phi$
between them, one can obtain DL if (1) $\eta$ is a zero of the Bessel function of order 0 and $\phi$ is an
odd
multiple of $\pi/2$ for equal and $\frac{\omega_1}{\omega_2}= odd~~integer$ driving frequencies, (2) $\eta$
is a zero of the Bessel function of order 0 and $\phi$ is an integer multiple of $\pi$ including zero for
$\frac{\omega_1}{\omega_2}= even~~integer \equiv m$, and (3)  $\phi = -\arcsin(\frac{J_0(\eta)}{\delta
J_m(\eta)})$ and $\eta$ is not a zero of the Bessel function of the even order $m$.
\end{abstract}

\maketitle
\nl

Needless to repeat the popular idea that the cold atoms in optical lattices provide a well controllable
experimental apparatus for testing the models of condensed matter physics\cite{exp}. Recently, in addition to
various other experimental verifications, the phenomenon of Dynamical Localization (DL) has been realized with
cold atoms in optical lattices\cite{lignier}. DL has been previously predicted in seminal work of Dunlap and
Kenkre\cite{kenkre}. DL states that the wave packet of a single particle moving in a single-band tight-binding
lattice with nearest-neighbor coupling driven by a spatially homogeneous ac field becomes localized whenever
the ratio of the  field magnitude to the field frequency is a root of the ordinary Bessel function of order 0.
This effect is latter understood physically in terms of dynamical band collapse\cite{holthaus} with far reaching
consequences including metal-insulator transitions in quasi-periodic lattices\cite{drese}. Dunlap-Kenkre result
has been recently generalized for arbitrary time-periodic forcing and going beyond nearest-neighbor
approximation\cite{dignam}. DL has also been predicted in semiconductor superlattices\cite{koch}. Recently the
importance of phase of driving field has been realized\cite{kudo} which show that, for systems with strong
attractive pairing, it enables different types of collisions and re-collisions between paired and un-paired
atoms. 

In this letter we point out the effect of phase difference between previously introduced off-diagonal
drive\cite{navinder} and diagonal drive\cite{kenkre} on the phenomenon of dynamical localization. We
analytically see that DL can be controlled with various other experimental control parameters and new DL
conditions exist. Our results are readily  amenable to experiments.
 
\begin{figure}
\includegraphics[height=5cm, width=8cm]{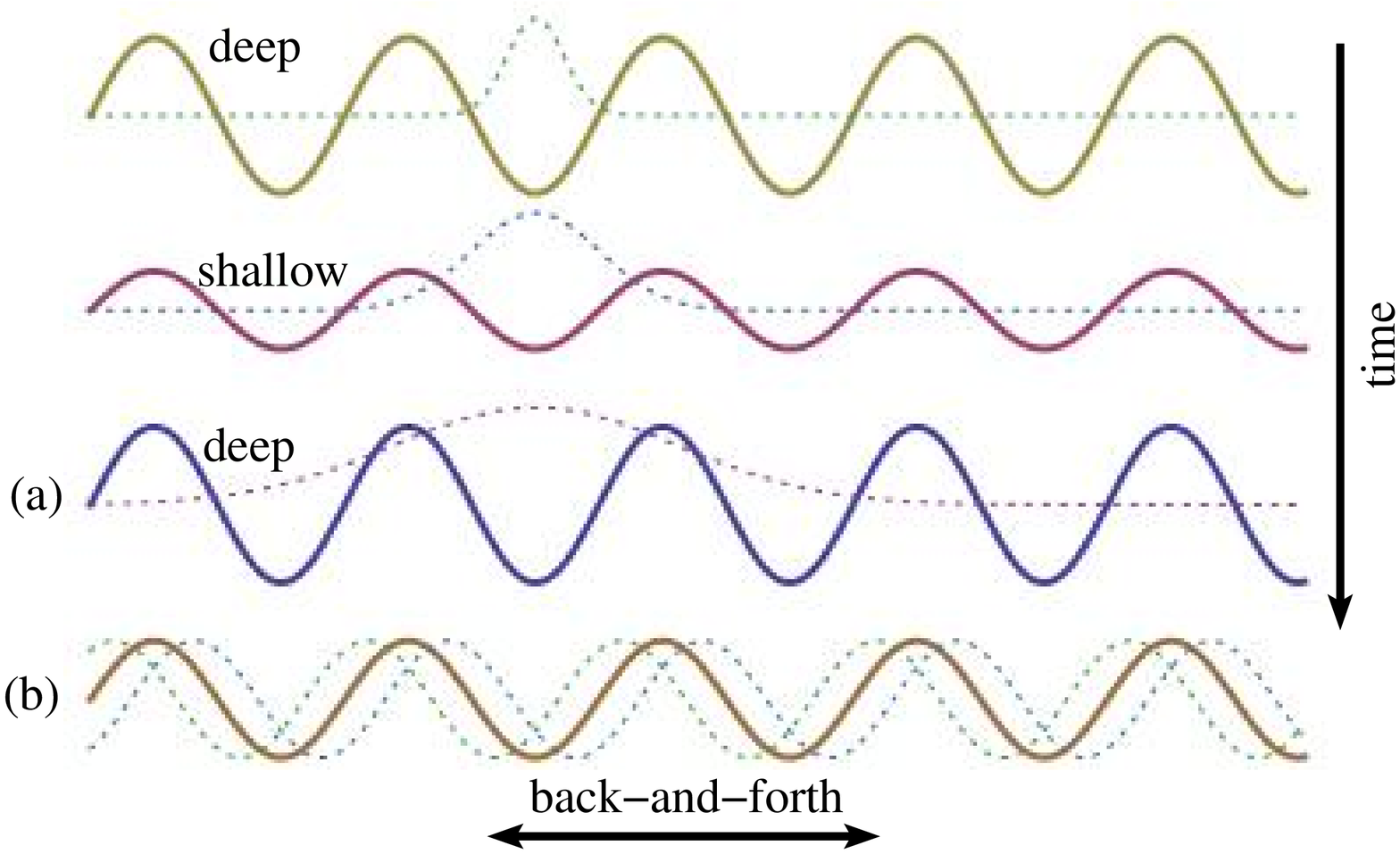}
\caption{Modulation of the optical lattice. The upper figure (a) corresponds to the
``deep-shallow-deep'' periodic modulation of the optical lattice (off-diagonal term in the
Hamiltonian with frequency $\omega_1$). The lower figure (b) represents the back-and-forth
motion (shaking) of the optical lattice with frequency $\omega_2$ (this corresponds to the
diagonal term in the Hamiltonian). The dotted line in the upper figure shows a schematic time evolution
of the atomic wave packet.}\label{optical}
\end{figure}

The Hamiltonian of an atom in an amplitude modulated and driven optical lattice (see Fig. \ref{optical}) in one
dimension is given as,
\begin{eqnarray}
\hat{H}(t) &=& -\frac{V}{2} (1+\delta \sin(\omega_1 t + \phi)) \sum_l (|l\rangle\langle l +1|
+ |l+1\rangle\langle l|)\nonumber\\ 
&& + \xi \cos(\omega_2 t) \sum_l l |l\rangle\langle l|.
\end{eqnarray}
Where $\xi$ and $\omega_2$ is the strength and frequency of the diagonal drive (equivalent to shaking the
optical lattice back and forth). $\delta$ and $\omega_1$ is the strength and
frequency of the off-diagonal modulation (by periodically modulating the amplitude of the optical
wells--``deep-shallow-deep'' periodic modulation). $\phi$ is the phase difference between the drives and $V$ is
the tunneling matrix element between the nearest neighbor optical wells. $|l\ra$ is the Wannier state localized
on lattice site $l$ (lattice constant $=1$). For $\delta = 0$ we have the case considered by Dunlap and Kenkre.

\begin{table*}[ht]
\caption{Summary of results}
  \begin{tabular}{|l||c|r|}
    \hline\hline
    $\omega_1 =\omega_2$ & $\frac{\omega_1}{\omega_2}=m$ (odd integer)  & $\frac{\omega_1}{\omega_2}=m$
(even integer) \\ \hline\hline
    $t^2 \frac{\beta^2}{2}(J_0^2(\eta) + \delta^2 \cos^2\phi J_1^2(\eta))$ & $t^2 \frac{\beta^2}{2}(J_0^2(\eta)
+ \delta^2 \cos^2\phi J_m^2(\eta))$ &  $t^2 \frac{\beta^2}{2}(J_0(\eta) + \delta \sin\phi J_m(\eta))^2$ \\
\hline
$\eta$ is a zero of $J_0$ and $\phi = (2 n+1)\frac{\pi}{2}$&$\eta$ is a zero of $J_0$ and
$\phi = (2 n+1)\frac{\pi}{2}$ & $\eta$ is a zero of $J_0$ and $\phi=n\pi$ \\
 (DL condition) & (DL condition) & or  $\phi =
-\sin^{-1}(\frac{J_0(\eta)}{\delta J_m(\eta)})$ and $\eta$ is not a zero of $J_m$. \\
    \hline
  \end{tabular}
\end{table*}


We start by putting the atom at a lattice site $0$ (Fig. \ref{optical}). Thus density matrix of the atom at
$t=0$ in site representation is $\rho_{m,n}(t=0) = \delta_{m,0}\delta_{n,0}$. The time evolution of the atomic
wave packet is given by Liouvelle-Von Neumann equation

\begin{equation}
i\hbar \frac{\pr \hat{\rho}(t)}{\pr t} = [\hat{H},\hat{\rho}].
\end{equation}

In the site representation it reads,
\begin{eqnarray}
\frac{\pr \la m|\hat{\rho}(t)|n\ra}{\pr t} &=& \frac{\pr \rho_{m,n}(t)}{\pr t} = \frac{i \beta}{2}
(1+\delta \sin(\omega_1 t + \phi))\nonumber\\
&&\times[\rho_{m+1,n} - \rho_{m,n+1} + \rho_{m-1,n} - \rho_{m,n-1}]\nonumber\\
&& - i \lambda \cos(\omega_2 t) (m-n)\rho_{m,n}(t).
\end{eqnarray}
Here $\beta \equiv V/\hbar,$ and$~\lambda\equiv \xi/\hbar$. 
Consider the following transformation
\begin{equation}
\rho_{m,n} = \bar{\rho}_{m,n} e^{- i f(t) (m-n)}, ~~~ f(t) = \frac{\lambda}{\omega_2} \sin(\omega_2 t).
\end{equation}

\begin{eqnarray}
&&\frac{\partial \bar{\rho}_{m,n}}{\partial t} = \frac{i \beta}{2} (1+\delta \sin(\omega_1 t +
\phi))\nonumber\\
&&\times(e^{-i f(t)}\bar{\rho}_{m+1,n}-e^{i f(t)}\bar{\rho}_{m,n+1}+e^{i
f(t)}\bar{\rho}_{m-1,n}\nonumber\\
&&-e^{-i f(t)}\bar{\rho}_{m,n-1}).
\end{eqnarray}
On writing the above equation in Fourier space with,
\begin{equation}
\tilde{\bar{\rho}}(k_1,k_2,t) = \sum_{m,n = - \infty}^{+\infty} \bar{\rho}(t) e^{- i m k_1} e^{i n k_2},
\end{equation}
puts it into a much simpler form,
\begin{eqnarray}
&&\frac{\partial \tilde{\bar{\rho}}(k_1,k_2,t)}{\partial t} = 2 i \beta (1+\delta \sin(\omega_1 t +
\phi))\nonumber\\
&&\times\sin(\frac{k_1+k_2}{2} -f(t))\sin(\frac{k_2-k_1}{2})\tilde{\bar{\rho}}(k_1,k_2,t).
\end{eqnarray}
This can be further simplified by defining center-of-mass and relative coordinates as $p \equiv
\frac{k_1+k_2}{2}$, and $u \equiv k_1 - k_2$, and re-defining $\tilde{\bar{\rho}}(k_1,k_2,t) \equiv
\varrho(p,u,t)$, whose solution is straightforward
\begin{equation}
\varrho(p,u,t)=e^{- 2 i\beta\sin(u/2) \int_0^t (1+\delta\sin(\omega_1 t' + \phi))\sin(p -
f(t')) dt'}.
\end{equation}
Where we have used $\varrho(p,u,0)= \sum_{m,n}\delta_{m,0}\delta_{n,0} e^{+i
(\lambda/\omega_2)\sin(\omega_2 t)(m-n)-i mk_1+in k_2}=1$. Now on integrating the above equation with respect to
$p$ we obtain a closed equation in $\xi(u,t)\equiv \frac{1}{2 \pi} \int_{-\pi}^{+\pi} \varrho(p,u,t) dp$.

We are interested in finding the mean displacement and mean-squared displacement of the atom from the starting
point. One notices that $\la n(t) \ra = i\frac{\pr \xi(u,t)}{\pr u}|_{u=0} = i \frac{1}{2\pi}\int_{-\pi}^\pi dp
\sum_{m,n} \bar{\rho}_{m,n}(t)\frac{\pr}{\pr u} e^{-i m (p+u/2) +i n (p-u/2)}|_{u=0} = \sum_n n \rho_{n,n}(t)$.
Similarly $\la n^2(t)\ra = -\frac{\pr^2 \xi(u,t)}{\pr u^2}|_{u=0}$.

A simple computation using the above prescription show that mean displacement is always zero (centre-of-mass of
the wave packet does not move). A computation of the mean-squared displacement leads to
\begin{equation}
 \la n^2(t)\ra = \frac{\beta^2}{2}(u^2(t)+v^2(t)),
\end{equation}
with $u(t) =\int_0^t dt' \cos(\eta\sin(\omega_2 t'))+\delta\int_0^t dt'\cos(\eta\sin(\omega_2
t'))\sin(\omega_1 t'+\phi),~~\eta=\frac{\lambda}{\omega_2}$ and $v(t) =\int_0^t dt' \sin(\eta\sin(\omega_2
t'))+\delta\int_0^t dt'\sin(\eta\sin(\omega_2 t'))\sin(\omega_1 t'+\phi)$. 

For the case of equal frequencies $\omega_1=\omega_2$ we write $u(t)$ and $v(t)$ in terms of time
bounded ($B_{u}(t)$ and $B_{v}(t)$) and time unbounded functions as $B_u(t) + t J_0(\eta)$ and $B_v(t) +
t \delta \cos(\phi) J_1(\eta)$. We obtain (after a long calculation) the mean-squared displacement in the long
time limit $t\gg\frac{2\pi}{\omega}$  as
\begin{equation}
 \la n^2(t)\ra = t^2 \frac{\beta^2}{2}(J_0(\eta)^2 + \delta^2 \cos^2\phi J_1(\eta)^2).
\end{equation}
Here $J_0$ and $J_1$ are the ordinary Bessel functions of order $0$ and $1$ and $\eta \ge 0$. For $\delta =0$
we get back the Dunlap-Kenkre result, as we should. Here we get DL if $\eta$ is the zero of $J_0$ and $\phi = (2
n+1)\frac{\pi}{2},~~n\in \mathcal{Z}$. One important implication is that the motion of the atom is now
controllable through $\delta$ and phase $\phi$ also.

For the case of unequal frequencies $\frac{\omega_1}{\omega_2} = m$ (integer) one sees that the analysis can be
further divided into two sub-cases (1) $m$ odd integer, and (2) $m$ even integer. As before, writing $u(t)$ and
$v(t)$ in terms of time bounded and time unbounded functions, we obtain $\la n^2(t)\ra$ for the first sub-case
($m$ odd integer) in the long time limit $t\gg\frac{2\pi}{\omega}$ as
\begin{equation}
 \la n^2(t)\ra = t^2 \frac{\beta^2}{2}(J_0(\eta)^2 + \delta^2 \cos^2\phi J_m(\eta)^2),
\end{equation}
and in the second sub-case ($m$ even integer) as
\begin{equation}
 \la n^2(t)\ra = t^2 \frac{\beta^2}{2}(J_0(\eta) + \delta \sin\phi J_m(\eta))^2.
\end{equation}
Here $J_m$ is the ordinary Bessel functions of order $m=\frac{\omega_1}{\omega_2}$ and $\eta \ge 0$. For $\delta
=0$ we again get back the Dunlap-Kenkre result as we should. Here we get DL (1) if $\eta$ is the zero of $J_0$
and $\phi = (2n+1)\frac{\pi}{2},~~n\in \mathcal{Z}$ (odd integer $m$ case),  (2) if $\eta$ is a zero of
$J_0$ and $\phi$ is an integer multiple of $\pi$ including zero (even integer $m$ case), and (3) if
$\phi_{critical} =
-\arcsin(\frac{J_0(\eta)}{\delta J_m(\eta)})$ and $\eta$ is not a zero of the Bessel function of the {\it even}
order $m$. These results are summarized in table I.

There is another remarkable dynamical result. If diagonal drive is zero i.e., $\xi =0$ i.e., $\eta =0$, the
off-diagonal drive
$\delta\sin(\omega_1 t+\phi)$ has {\it no effect} on the temporal evolution of the atomic wave packet. The width
of the wave packet expand $\propto t^2$ as it should in pure quantum dynamics (it will not see the
``deep-shallow-deep'' motion of the optical lattice !).

The above mathematical results should be directly verifiable with present sophistication of experiments
with cold atoms in optical lattices\cite{lignier} especially the $\phi_{critical} =
-\sin^{-1}(\frac{J_0(\eta)}{\delta J_m(\eta)})$.


\begin{thebibliography}{100}

\bibitem{exp} D. Jaksch and P. Zoller, Ann. Phys. {\bf 315}, 52 (2005); S. Giorgini, L. P. Pitaevskii,
and S. Stringari, Rev. Mod. Phys. {\bf 80}, 1215 (2008); I. Bloch, J.  Dalibard, and W. Zwerger, Rev. Mod. Phys.
{\bf 80}, 885 (2008); O. Morsch and M. Oberthaler, Rev. Mod. Phys. {\bf 78}, 179 (2006); A. Polkovnikov,
K. Sengupta, A. Silva, and M. Vengalattore, Rev. Mod. Phys. 0{\bf 83}, 863 (2011); A.R. Kolovsky and H.J.
Korsch, Inter. J. Mod. Phys. B. {\bf 18} 1235 (2004); M. Raizen, C. Salomon, and Q.  Niu, Phys. Today, page 30,
July 1997. 



\bibitem{lignier} H. Lignier, C. Sias, D. Campini, Y. Singh, A. Zenesini, O. Morsch, and E.
Arimondo, Phys. Rev. Lett. {\bf 99}, 220403 (2007); A. Zenesini1, H. Lignier1, D. Ciampini, O. Morsch,
and E. Arimondo, Phys. Rev. Lett. {\bf 102}, 100403 (2009); A. Eckardt, M. Holthaus, H. Lignier,
A. Zenesini, D. Ciampini, O. Morsch, and E. Arimondo, Phys. Rev. A {\bf 79}, 013611 (2009); K. W. Madison, M. C.
Fischer, R. B. Diener, Qian Niu, and M. G. Raizen, Phys. Rev. Lett. {\bf 81}, 5093 (1998). 



\bibitem{kenkre} D. H. Dunlup and V. M. Kenkre, Phys. Rev. B. {\bf 34}, 3625 (1986).
%


\bibitem{holthaus}M. Holthaus, Phys. Rev. Lett. {\bf 69}, 351 (1992); M. Holthaus and D. Hone, Phys. Rev. B.
{\bf 47}, 6499 (1993).


\bibitem{drese} K. Drese and M. Holthaus, Chem. Phys. {\bf 217}, 201 (1997); K. Drese and M. Holthaus, Phys.
Rev. Lett. {\bf 78}, 2932 (1997);  D. J. Bores, B. Goedeke, D. Hinrichs, and M. Holthaus, Phys. Rev. A. {\bf
75}, 063404 (2007).



\bibitem{dignam} M. M. Dignam, C. Martijn de Sterke, Phys. Rev. Lett. {\bf 88}, 046806 (2002).


\bibitem{koch}T. Meier, G. von Plessen, P. Thomas, and S. W. Koch, Phys. Rev. B, {\bf 51}, 14490 (1995).


\bibitem{kudo} K. Kudo and T. S. Monteiro, arXiv: 1008.2096 (2010).


\bibitem{navinder} N. Singh, J. Phys. A: Math. Theor. {\bf 41}, 255001 (2008).



%
%

\end{thebibliography}
\end{document}